\documentclass[twocolumn, prb]{revtex4}
\usepackage{graphicx}
\usepackage{dcolumn}
\usepackage{amsmath}
\begin{document}
\hsize\textwidth\columnwidth\hsize\csname@twocolumnfalse\endcsname
\title{Correlated charge detection for read-out of a solid state quantum computer}
\author{T.  M.  Buehler,$^{1,2}$ D. J. Reilly,$^{1,2}$ R. Brenner,$^{1,2}$ A.  R.  Hamilton,$^{1,2}$
A.  S.  Dzurak,$^{1,3}$ R.  G.  Clark$^{1,2}$}
\affiliation{$^1$Centre for Quantum Computer Technology, University of New South Wales,
Sydney 2052, Australia} \affiliation{$^2$School of Physics, University of New South
Wales, Sydney 2052, Australia} \affiliation{$^3$School of Electrical Engineering \&
Telecommunications, University of New South Wales, Sydney 2052, Australia}

\begin{abstract}
\noindent The single electron transistor (SET) is a prime candidate for reading out the
final state of a qubit in a solid state quantum computer. Such a measurement requires
the detection of sub-electron charge motion in the presence of random 
charging events. We present a detection scheme where the signals from 
two SETs are cross-correlated to suppress unwanted artifacts due to charge 
noise.   This technique is demonstrated by using the two SETs to detect the 
charge state of two tunnel junction - coupled metal dots, thereby simulating 
charge transfer and readout in a two qubit system.   These measurements
indicate that for comparable buried dopant semiconductor architectures the 
minimum measurement time required to distinguish between the two charge
states is of the order of 10 ns.

\end{abstract}

\maketitle

Quantum computers (QCs) hold promise to unveil unprecedented computational power by exploiting
quantum mechanical principles such as superposition and entanglement
\cite{preskill_qcreview}. Of the many proposals to implement such machines, architectures
based on the manipulation of quantum two-level systems (qubits) in the solid state are
appealing due to their scalability and ease of integration with existing
micro-electronics hardware \cite{Kane_nature,supercon_qubits,vrijen}. However, the
coupling of solid state qubits to a classical system for the purpose of read-out remains
a formidable challenge, amounting to the detection of single charge or spin states in
the presence of background artifacts. Several recent proposals for solid state QCs
make use of the extreme charge sensitivity of single electron
transistors (SET) to infer the state of a qubit either via direct detection \cite{divincenzo,schoelkopf_nature} or by 
sensing a spin-dependent charge transfer event \cite{Kane_nature,vrijen}. Since in these cases 
the read-out signal is indeterministic, read-out schemes
require good signal to noise without relying on the use of frequency 
selective (lock-in) or auto-correlation noise reduction methods.

Presently, the dominant decoherence mechanism for charge based qubits is $1/f$ 
background charge noise \cite{nakamura2}. In order to overcome this 
limitation, qubit operations should occur at speeds well above the $1/f$ 
corner (3kHz), where the influence of background charge motion is 
reduced. The process of read-out however, is stochastic in nature and can 
potentially be affected by slow time-scale random-charging events close to or within the 
SETs.   Such charging events are indistinguishable from real charge {\it signals} connected with read-out and constitute
a strong challenge for single-shot projective measurements.

Here we focus on a read-out detection technique where the independent charge signals 
from two aluminum SETs are cross-correlated  both temporally and spatially in order to greatly reduce
the effect of spurious background charge noise.  Previously 
Zorin {\it et al.,} \cite{zorin} have made use of two SETs to identify and 
spatially resolve different {\it sources} of intrinsic charge noise associated with 
$1/f$ type trapping processes\cite{weissman}. Here we use a similar 
technique not to discriminate between different noise sources, but as a means 
to distinguish the real {\it signal} from background artifacts with the 
fidelity required to read-out a quantum computer.  

\begin{figure}
\begin{center}
\includegraphics[width=8cm]{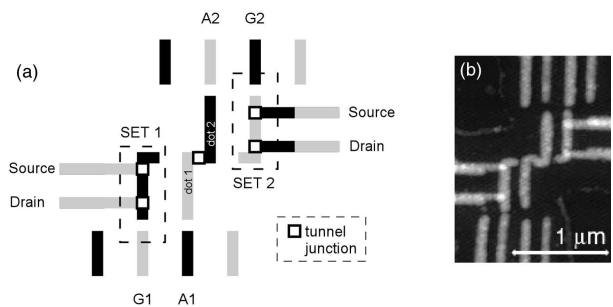}
\caption{(a) Schematic and (b) AFM image of a twin-SET device, fabricated by
double-angle evaporation of Al with intermediate oxidation to form tunnel junctions.}
\vspace{-0.5cm}
\end{center}
\end{figure}

To demonstrate an ability to detect a fraction of an 
electron charge, as required for qubit read-out, we have designed and fabricated an all-metal
twin-SET/double-dot system (Figure 1). In this design two metal dots connected by a
tunnel junction \cite{amlani} simulate a solid state charge qubit. When an electron tunnels
from dot1 to dot2, SET1 senses the departure of the electron, while 
SET2 simultaneously senses the arrival of the electron.  Cross-correlation of the two signals permits suppression of all charging events except
those that originate from the area in between the two SET detectors.

Devices were fabricated using electron beam
lithography and a bi-layer resist process \cite{dolan} on a phosphorus 
doped silicon substrate ($10 \Omega cm$). The Al/AlO$_x$ tunnel junctions
were formed by standard shadow evaporation processes with an intermediate oxidation
step. A thin native oxide layer was used in conjunction with the P-doped 
substrate to produce additional charge traps for the purpose of
characterising our cross-correlation detection scheme.  Note that although
our device is engineered to contain additional charge traps, devices
fabricated on high quality substrates are also influenced by background charge noise \cite{zimmerman}.
Measurements
were performed in a dilution refrigerator with a base temperature below 30 mK, using
standard lock-in techniques with excitation voltages of $10-50 \:\mu V$.
A magnetic field of 0.5 Tesla was applied to suppress superconducting effects in 
the aluminum nanostructures.  

\begin{figure}
\begin{center}
\includegraphics[width=8.0cm]{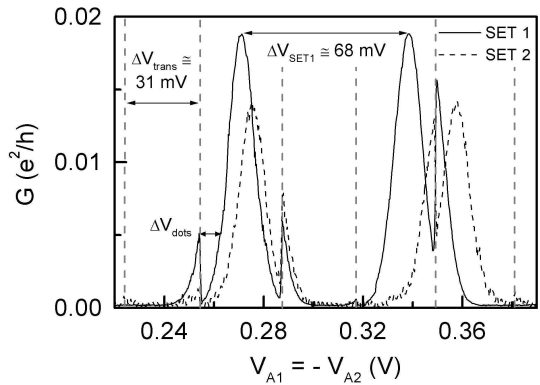}
\caption{Conductance of the two SETs as a function of differential A-gate 
bias $V_{A1}=-V_{A2}$.     Electron transfer between the dots is seen as periodic `jumps' in
the conductance of both SETs as highlighted by the vertical dashed lines.  }
\vspace{-0.5cm}
\end{center}
\end{figure}

This device simulates read-out in a solid state quantum computer by using SETs 
to sense the controlled transfer of an electron from one metal dot to the 
other.  Controlled electron transfer is achieved by differentially biasing 
gates A1 and A2 to create an electric field between the dots. As the field is increased, the potential difference 
between the dots becomes large enough for one electron to overcome the 
Coulomb charging energy of 
the double-dot system. Figure 2 shows the conductance of both SET1 and SET2 as a function 
of the differential A-gate bias.   The bias 
gates induce conventional Coulomb blockade (CB) oscillations in both SETs as the electrostatic potential of the SET islands is varied.  
Superimposed on these oscillations are abrupt discontinuities or `jumps' associated 
with charge transfer between the metal dots (vertical dashed lines in Fig.  2). The 
magnitude of these jumps depends on the transconductance of the SET.  The combination of 
double-dot electron transfer and CB oscillations produces signals with two distinct 
periods: A differential gate bias of 
$\Delta V_{SET1} \simeq 68$ mV for SET1 and $\Delta V_{SET2} \simeq 82.5$ mV for SET2 corresponds to the 
addition of one electron onto the SET island, while a differential bias of 
$\Delta V_{trans} \simeq 31$mV is required to transfer an electron between the 
metal dots. 

\begin{figure}
\begin{center}
\includegraphics[width=8.0cm]{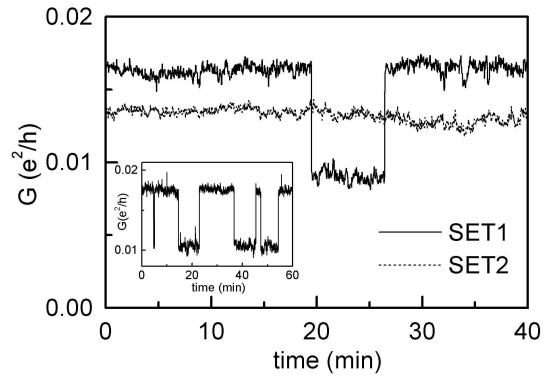}
\caption{Conductance of the two SETs as a function of time.   The inset 
shows random telegraph signals (RTSs) associated with a charge trap close 
to SET1. The body of the figure shows a RTS occurring only in SET1 with the 
conductance of SET2 remaining constant.}

\vspace{-0.5cm}
\end{center}
\end{figure}

Having demonstrated the ability to detect controlled single 
electron transfer in the double-dot system, we now turn to the issue of 
charge noise.   Figure 3 
presents data 
taken on both SETs as  a function of {\it time}, 
with all gate biases fixed.   
The data shown in the inset to Fig. 3 reveals the presence of random telegraph signals (RTSs)
associated with the charging - decharging of traps in the vicinity of the  SET island \cite{amarasinghe}. Note
that for the data shown in the main body of the figure the 
conductance of SET2 remains constant and free of switching signals for the
duration of the measurement. Taken on its own a single RTS signal from SET1 would be
inseparable from a true charge event associated with the transfer of an electron on the
double-dot system. However, by correlating the signals from {\it both} SETs 
we are able to distinguish between readout events and background charge 
noise.

\begin{figure}
\begin{center}
\includegraphics[width=8.0cm]{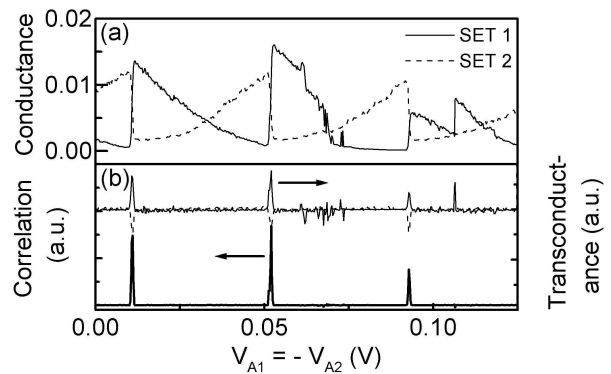}
\caption{\textbf{a)} Conductance of the two SETs as a function of 
differential A-gate bias, with direct coupling of V$_{A1}$ and V$_{A2}$ to the 
SETs removed by a compensating bias on G1 and G2. The characteristic sawtooth behavior results from controlled 
electron transfer between the metallic dots.   
\textbf{b)} Transconductance ($\partial G/\partial V_A$) and cross-correlation  of the two SET signals to distinguish 
charge transfer events from background charge fluctuations. }
\vspace{-0.5cm}
\end{center}
\end{figure}

For efficient read-out the SETs should be operated 
at their maximum sensitivity, half way up a CB peak where the 
transconductance is largest.   In order to achieve this while 
simultaneously continuing A-gate operations, the SET gate biases (G1 and 
G2) must be adjusted to compensate for the changing A-gate potential. 
Figure 4(a) shows data for both SETs where the direct coupling of the 
A-gate to the SETs has been compensated by adjusting gates G1 and G2. In 
this case the data for both SETs shows a clear periodic sawtooth,
corresponding to the transfer of single electrons between the metal dots.
If both SETs are biased on the same side of a Coulomb blockade peak (eg on
the rising edge) then their compensated signals move in {\it opposite} directions when a
transfer event occurs.  This is because SET1 senses the  departure of
an electron and SET2 senses its arrival. In contrast,
signals that do not exhibit this behavior can be rejected as background
charge noise.   Figure 4(a) shows such an event occurring near
$V_{A1}\approx 0.11V$. By correlating the signals from both SETs it is 
possible to reject all events except those signals
arising from electron transfer between the metal dots.     Figure 4(b)
shows the  correlation of signals by multiplying the transconductances
($\partial G_{SET1}/\partial V \times \partial G_{SET2}/\partial V$). A clear
suppression of random charging events is achieved.

For SETs to be used to infer the state of qubits in QC architectures, the 
time required to perform a measurement must be less than the mixing time of 
the two level system (qubit).   The ultimate limit to the SET's sensitivity 
is due to shot noise \cite{korotkov}, which provides a lower bound for the time required to 
measure a charge $\Delta q$ of 
$(t_{min})^{1/2} = 1.2 \times 10^{-6}e/\Delta q $.  Therefore the amount  
of charge $\Delta q$ induced on the SET island is a critical parameter characterising 
the efficiency of the read-out process, as it determines the time $t_{min}$ 
required to distinguish the desired signal from the shot noise.   For the 
simulation device presented here, the change in induced charge $\Delta q$ 
 due to the transfer of a single electron between the 
double dots can be determined from Fig. 2. We infer 
$\Delta q$ by relating the lateral shift of the conductance $G$ in 
A-gate bias due to a charging event, $\Delta V_{dot} \approx $8.1mV (SET1) and
$\Delta V_{dot} \approx $7.1mV (SET2), to the 
gate bias required to add one extra electron to the SET island, 
$\Delta V_{SET1} \approx$ 68mV and $\Delta V_{SET2} \approx$ 82.5mV. 
For the data from SET1 in Fig. 2,   
$\Delta q$ is estimated to be $\approx$8.1mV/68mV $\rightarrow \Delta q$ = 
0.12e (for SET2 $\Delta q$ = 0.086e).  
Therefore a SET with a sensitivity approaching the quantum limit, would 
require a minimum measurement time $t_{min}$ of order 10 ns to 
distinguish a single charge transfer event in the double-dot system from 
the shot noise, with a signal to noise ratio of 10 ($\Delta q = 0.01e$).

The minimum 
measurement time places significant practical constraints on solid state QC 
architectures, since the qubit to be read out with the SET must remain 
constant during the read-out process.   For architectures utilising single 
dopants in a semiconductor, such as P in Si\cite{Kane_nature}, numerical modeling\cite{green} indicates 
that the capacitive coupling $\kappa$ between the dopant and the SET is 
similar to that in our metal nanostructure, suggesting that the readout times 
will be comparable for the two cases.

In conclusion, we have demonstrated a charge detection scheme that makes use of cross -
correlated signals from two SETs to suppress spurious noise associated with random
trapping-detrapping events in the substrate and oxide. This scheme was demonstrated
using an all metal twin-SET double-dot system to distinguish controlled electron
transfer from random telegraph signals originating close to the SETs, and simulates
readout in a solid state two-qubit system. Our measurements indicate that for comparative
capacitive coupling,  readout using a quantum-limited SET requires 
$\approx$10 ns
to distinguish the readout signal with good signal to noise. Work is 
presently underway to construct twin radio frequency SET readout devices, so that this 
correlated measurement technique can be extended to much shorter 
time-scales than presently demonstrated.   We are also developing twin-SET devices in which the two 
metal dots are replaced with two clusters of localised P dopants buried in 
the silicon substrate, as a step towards the goal of creating a scalable P 
in Si quantum computer.

We thank R.  P.  Starrett and  D. Barber for technical support and F. Green for informative
discussions. This work was supported by the  Australian Research Council,
the Australian Government, by the US National Security Agency (NSA),
Advanced Research and Development Activity (ARDA) and the Army Research
Office (ARO) under contract number DAAD19-01-1-0653. DJR acknowledges a
Hewlett-Packard Fellowship.

\end{document}